\documentstyle[preprint,revtex]{aps}
\tightenlines
\begin{document}
\preprint{HD--THEP--92--32}
\draft
%
%
\begin{title}
{\bf Note on Tests of the Factorization Hypothesis and the\\
Determination of Meson Decay Constants}
\end{title}
\author{Volker Rieckert}
\begin{instit}
Institut f\"ur Theoretische Physik der Universit\"at Heidelberg\\
Philosophenweg 16, D-6900 Heidelberg, Germany\\
e-mail:  K46 @ DHDURZ1
\end{instit}
%
%
\begin{abstract}
We discuss various tests of the factorization hypothesis making use
of the close relationship between semi-leptonic and factorized
nonleptonic decay amplitudes.  It is pointed out that factorization
leads to truely model-independent predictions for the ratio of
nonleptonic to semi-leptonic decay rates, if in the nonleptonic decay
a spin one meson of arbitrary mass or a pion take the place of the
lepton pair.  Where the decay constants of those mesons are known,
these predictions represent ideal tests of the factorization
hypothesis.  In other cases they may be used to extract the decay
constants.  Currently available data on the decays $\bar B^0 \to
D^+\pi^-,\, D^{*+}\pi^-,\, D^+\varrho^-,\, D^{*+}\varrho^-$ are shown
to be in excellent agreement with the factorization results.  A
weighted average of the four independent values for the QCD
coefficient $a_1$ extracted from the data gives $a_1=1.15\pm 0.06$
suggesting that it may be equal to the Wilson coefficient $c_1(\mu)$
evaluated at the scale $\mu = m_b$.
\end{abstract}
%
%
\narrowtext
The dynamics of nonleptonic weak decays is strongly influenced by the
confining color forces among the quarks.  In contrast to
semi-leptonic transitions, where the lepton current naturally
factorizes and one is left with the hadronic matrix element of a
color-singlet quark current, nonleptonic processes are complicated by
the phenomenon of quark rearrangement due to the exchange of soft and
hard gluons.  The theoretical description involves matrix elements of
local four-quark operators, which are much harder to deal with than
current operators.

A great simplification can be accomplished if one is willing to adopt
the factorization hypothesis, which relates the complicated
nonleptonic decay amplitudes to products of meson decay constants and
hadronic matrix elements of current operators similar to the ones
encountered in semi-leptonic decays.  Despite its remarkable success
in the description of 2-body decays of $B$- and $D$-mesons, precise
tests of the factorization hypothesis are of utmost importance in
order to find out its realm of applicability as well as its
limitations.  While many tests have been suggested or already carried
out~\cite{Bj89,rosner,bo90,mannelD,mannelB,buch,pham,reader}, most of
them do not simply test the factorization hypothesis, but rather
factorization together with some phenomenological model or,
alternatively, together with heavy-quark symmetry for dealing with
the hadronic current matrix elements.  It is the main objective of
this short note to concentrate on such tests that do not suffer from
additional uncertainties due to our unsatisfactory ways of dealing
with non-perturbative QCD.

As there exist several versions of factorization in the literature,
let us begin by giving an unambiguous prescription of how to
calculate the rate of some exclusive nonleptonic $B$-decay in the
factorization approximation.  We will concentrate on $b \to c$
transitions, which are induced by the effective Hamiltonian
\begin{eqnarray}
H_{eff} & = &{G_F \over \sqrt{2}}\,V_{cb}\,\left[ c_1(\mu)\,Q_1^{cb}
                                      +c_2(\mu)\,Q_2^{cb}\right]
\nonumber
\\[3pt]
        & + &{\rm penguin~operators}.
\label{Heff}
\end{eqnarray}
It consists of products of local four-quark operators with
scale-dependent Wilson coefficients $c_i(\mu)$.  The operators $Q_1$
and $Q_2$, written as products of color-singlet currents, are given
by
\begin{eqnarray}
Q_1^{cb} & = & \left[ (\bar d'u)_{_{V-A}}\,+\,(\bar s'c)_{_{V-A}}
              \right]
(\bar c\,b)_{_{V-A}}~, \nonumber \\[3pt]
Q_2^{cb} & = & (\bar c\,u)_{_{V-A}}\,(\bar d'b)_{_{V-A}}\;+\;
(\bar c\,c)_{_{V-A}}\,(\bar s'b)_{_{V-A}}~,
\label{Obc}
\end{eqnarray}
where $d'$ and $s'$ denote weak eigenstates of the down and strange
quarks,
respectively, and $(\bar c\,b)_{_{V-A}}=\bar
c\,\gamma_{\mu}\,(1-\gamma_5)\,b$
etc. The Wilson coefficients of so-called penguin
operators~\cite{Sh77} in
Eq.~(\ref{Heff}) are very small.  Their contribution to the dominant
decay amplitudes may be neglected.

If QCD was turned off, the Wilson coefficients of the operators
$Q_1^{cb}$ and $Q_2^{cb}$ would be $c_1 = 1$ and $c_2=0$.  These
values are modified by hard gluon exchange.  Evaluated at the scale
$\mu=m_b\simeq 5.0$~GeV one finds in leading logarithmic
approximation~\cite{Al74}:  $c_1(m_b)=1.12$ and $c_2(m_b)=-0.26$.

According to the factorization hypothesis one may now write the
hadronic matrix elements of $Q_1^{cb}$ and $Q_2^{cb}$ as products of
two current matrix elements~\cite{factorization}.  As an example, we
consider the decay amplitude of the transition  $\bar B^0\to
D^+\,\pi^-$, which in the factorization approximation is given by
\begin{equation}
 A^{fact}={G_F\over \sqrt{2}}\,V_{cb}\,V_{ud}^*\,a_1\,
          \langle\,\pi^-\,\vert\:(\bar d\,u)_A\,\vert\,0\,\rangle\:
          \langle\,D^+\,\vert\:(\bar c\,b)_V\,\vert\,
	  \bar B^0\,\rangle~.
\label{Afact}
\end{equation}
Class I transitions like the one considered above, in which only a
charged meson can be generated directly from a current, are
proportional to the QCD coefficient $a_1$.  Its relation to the
Wilson coefficients will be discussed below.  Correspondingly, those
decays in which the meson generated directly from the current is
neutral, like the $J/\Psi$-particle in the decay
$\bar B \to \bar K\,J/\Psi$, are called class II, and their decay
amplitudes are
proportional to the QCD-coefficient $a_2$.  Factorized amplitudes in
which there is interference between $a_1$- and $a_2$-terms are
categorized as class III.

Usually, form factor suppressed weak annihilation topologies
($W$-exchange and  quark-annihilation diagrams) are neglected in the
calculation of factorization amplitudes.  This is not an inherent
property of the factorization approximation.  Rather it is necessary
from a practical point of view, since little more is known about form
factors at such large time-like momentum transfer than that they
should be strongly suppressed.  What really is an inherent property
of the factorization approximation is the neglect of final state
interactions (FSI).  However, unlike $D$-decays, the decays of
$B$-mesons do not take place in a resonance region.  Thus one has
good reason to believe that ignoring the effects of FSI is a good
approximation in $B$-decays.

Let us now turn to the relation between the Wilson coefficients and
the QCD coefficients $a_1$ and $a_2$.  Naively, one would expect
$a_1=c_1(\mu_f) + \xi c_2(\mu_f)$ and $a_2=c_2(\mu_f) + \xi
c_1(\mu_f)$, with $\xi=1/N_c$, and $\mu_f$ denoting the factorization
point, in $B$-decays usually identified with $m_b$.  However,
experience in $D$-decays has shown that setting $\xi =0$ allows for a
better description of the data, and it has been suggested to treat
$\xi$ or even $a_1$ and $a_2$ independently as a free
parameters~\cite{bsw87}.  Thus one can test the factorization
hypothesis by checking whether or not the values for the QCD
coefficient $a_1$ ($a_2$) as extracted from different class I (class
II) transitions agree with each other.  For $a_1$ also an absolute
prediction becomes possible, by observing that varying the parameter
$\xi$ in the range $0<\xi <1/3$ induces no more than a 10\% change in
$a_1$.  One would therefore expect $a_1 = 1.1\pm 0.1$ which has been
confirmed in a recent extraction of $a_1$ from all available
nonleptonic 2-body decays of $B$-mesons~\cite{buch}.  However, it
should be stressed that the fit for $a_1$ has been strongly dominated
by the two decay modes\footnote{Here and in the following,
``$D^{(*)}$'' stands for ``$D$ or $D^*$''} $\bar B^0 \to D^{(*)+}\,
\pi^-$.  It remains to be seen whether the same value of $a_1$ will
be found from other decay modes as more precise data become
available.

As first pointed out by Bjorken, the close relationship between
factorized amplitudes and semi-leptonic decay amplitudes provides the
most direct test of the factorization
assumption~\cite{Bj89,rosner,bo90}.  To this end, a nonleptonic decay
width is related to the corresponding differential  semi-leptonic
decay width evaluated at the same $q^2$.  Let us consider the ratios
\begin{equation}
   R_P^{(*)}
   =\frac{\Gamma (\bar B^0\to D^{(*)+}\,P^-)}
        {\left. {\rm d} \left\{ \Gamma (\bar B^0\to
	D^{(*)+}\,\ell^-\,\bar
         \nu_{\ell})\right\}/{\rm d} q^2\,\right| _{q^2=m_P^2}}
   =6\pi^2\,f_P^2\,\vert a_1\vert ^2\,\vert V_{ij}\vert^2 \,
    X_P^{(*)}~,
\label{RP}
\end{equation}
where $f_P$ is the decay constant of the pseudoscalar meson $P$,
$V_{ij}$ is the appropriate KM-matrix element (associated with $P$)
and (in the limit of vanishing lepton mass)
\begin{eqnarray}
  X_P &=&
  \frac{(m_B^2-m_D^2)^2}
       {\left[ m_B^2-(m_D+m_P)^2\right]\,\left[
       m_B^2-(m_D-m_P)^2\right]} \,
  \left| {F_0(m_P^2)\over F_1(m_P^2)}\right| ^2~,\nonumber\\[3pt]
  X_P^* &=&
  \left[ m_B^2-(m_{D^*}+m_P)^2\right]\,\left[
 m_B^2-(m_{D^*}-m_P)^2\right]\,
  \frac{\vert A_0(m_P^2)\vert^2}
       {m_P^2\,{{\displaystyle} \sum_{i=0,\pm}} \vert
       H_i(m_P^2)\vert^2}~.
\label{XP}
\end{eqnarray}
The helicity amplitudes $H_0(q^2)$ and $H_{\pm}(q^2)$ are defined in
Ref.~\cite{Ba/Wi}.

Bjorken has suggested this test with $P=\pi$, in which case
$X_{\pi}\simeq X^*_{\pi}\simeq 1$ to within less than $0.5\%$ as can
be shown by expanding those quantities in powers of
$m_{\pi}^2/m_B^2$~\cite{buch}.  For heavier pseudoscalar mesons,
$X_P$ and $X_P^*$ become model dependent and may quite substantially
deviate from 1.  In the infinite quark mass limit, one finds, for
example, $X_{D_s}\simeq 1.36$ and $X_{D_s}^*\simeq 0.37$~\cite{buch}.

We can get rid of this model dependence altogether by replacing the
pseudoscalar meson $P$ in Eq.~(\ref{RP}) by a vector or pseudovector
meson.  In the factorization approximation one then finds
\begin{equation}
   R_V^{(*)}
   =\frac{\Gamma (\bar B^0\to D^{(*)+}\,V^-)}
        {\left. {\rm d} \left\{ \Gamma (\bar B^0\to
	D^{(*)+}\,\ell^-\,\bar
         \nu_{\ell})\right\}/{\rm d} q^2\,\right| _{q^2=m_V^2}}
   =6\pi^2\,f_V^2\,\vert a_1\vert ^2\,\vert V_{ij}\vert^2~,
\label{RV}
\end{equation}
where now, of course, the KM-matrix element is associated with the
(pseudo-)vector meson and $f_V$ denotes its decay constant.  The
reason for all form factors and additional kinematical factors to
cancel in the ratio can be easily understood.  For zero lepton
masses, the lepton pair that in the semi-leptonic decay is generated
by the $(V-A)$ current carries spin one in its c.m.~frame.
Integrated over the lepton angles keeping
$q^{\mu}=(p_{\ell}+p_{\nu})^{\mu}$ fixed, the production of the
lepton pair is therefore kinematically equivalent to the production
of a (pseudo-)vector particle with four-momentum $q^{\mu}$ (summed
over all polarizations of the (pseudo-)vector particle).  Corrections
to Eq.~(\ref{RV}) due to finite lepton masses are of order
$m_{\ell}^2/m_V^2$.  With the $\varrho$-meson being the lightest
spin-one meson these corrections may safely be neglected for
electrons and muons.

Setting $V=\varrho$, we can use Eq.~(\ref{RV}) to obtain two
independent values for $a_1$, since the decay constant $f_{\varrho}$
is known\footnote{We use $f_{\varrho}=205$~MeV~\protect\cite{buch}}.
These values should be compared with those obtained from
Eq.~(\ref{RP}) with $P=\pi$.  However, as long as the differential
$q^2$ spectrum of the semi-leptonic decay $\bar B^0 \to
D^+\,\ell^-\,\bar\nu$ has not yet been measured, we must again resort
to some form factor model in decays with a $D$-meson in the final
state.  In Table~\ref{tab:ratios}, we have used the predictions of
Ref.~\cite{buch} for those two decays.  They are based on an
Isgur-Wise function extracted from data on the decay $\bar B^0 \to
D^{*+}\,\ell^-\,\bar\nu$ with perturbative QCD-corrections and
(model-dependent) $1/m_Q$-corrections added on.  Nonleptonic decay
data used in Table~\ref{tab:ratios} has been taken from
CLEO~\cite{kh} and ARGUS~\cite{argus}.  The ARGUS data as well as the
predictions of Ref.~\cite{buch} have been rescaled using the new
CLEO measurement BR$(D^{*+} \to D^0\,\pi^+)=(68\pm 2)\%$~\cite{cleo}.
The experimental number for the ratio $R_{\varrho}^*/R_{\pi}^*$, where
some of the systematic uncertainties drop out, has been taken from
CLEO alone.  We observe that all four values for the QCD coefficient
$a_1$ presented in Table~\ref{tab:ratios} are in excellent agreement
with each other and with the expectation from perturbative QCD, thus
providing strong support for the factorization hypothesis in
$B$-decays with large recoil.  Taking the weighted average of all
four values gives $a_1=1.15\pm 0.06$, which suggests that, just like
in $D$-decays, we may have $\xi =0$, i.e.\ the QCD coefficient $a_1$
may be equal to the Wilson coefficient $c_1(\mu)$ evaluated at the
scale of the decaying quark.

As better statistics becomes available the decays $\bar B^0\to
D^{(*)+}\,K^{(*)-}$ should be included in the above analysis, since
the decay constants $f_K$ and $f_{K^*}$ are also known (the latter
can be extracted from exclusive $\tau$-decay data).
On the other hand, Eq.~(\ref{RV}) may be used together with the
experimentally determined value of the QCD coefficient $a_1$ to
extract yet unknown decay constants of spin-one mesons like the
$a_1$-meson or the $D_s^*$-meson without resorting to some particular
form factor model or to heavy-quark symmetry.

{}From our kinematical argument about the equivalence of the lepton
pair in the semi-leptonic decay and the spin-one particle in the
nonleptonic decay it is clear that Eq.~(\ref{RV}) must be valid,
separately, for the different polarizations of the $D^*$-meson in the
final state.  This amounts to the factorization prediction that the
polarization of the $D^*$-meson in the nonleptonic decay $\bar B^0
\to D^{*+}\,V^-$ should be equal to the polarization in the
corresponding semi-leptonic decay at the same $q^2$.  This prediction
is currently being tested by the CLEO collaboration~\cite{kh}.
However, in interpreting the results of such a test, one has to bear
in mind that in the semi-leptonic as well as in the nonleptonic case
the $D^*$-polarization at the points $q^2=0$ and $q^2=q^2_{\rm max}$
is unambiguously determined by kinematics alone to be 100\%
longitudinal and $1/3$ longitudinal, respectively.  At zero recoil,
there is no preferred direction and thus the value $1/3$ just
expresses the fact that there are two transverse, but only one
longitudinal polarization.  At $q^2=0$, corresponding to maximum
recoil in the semi-leptonic decay, the left-handed electron or muon
and the right-handed anti-neutrino go off parallel to each other,
thereby forcing the $D^*$ into longitudinal polarization.  In the
corresponding nonleptonic decay $\bar B^0 \to D^*\,V^-$ we know (even
without the factorization approximation!) that the decay amplitude
must be proportional to the polarization vector of the
(pseudo-)vector meson $V$.  Now, for small $q^2=m_V^2\ll m_B^2/4$ the
(pseudo-)vector meson $V$ is highly relativistic (in the $B$-meson
rest frame) so that for longitudinal polarization of $V$ (and
consequently $D^*$) the components of the polarization vector acquire
very large values, causing longitudinal polarization to dominate.

The above discussion shows that in comparing polarizations in
semi-leptonic and nonleptonic decays, one needs polarization data of
quite high precision in order to make a statement about the validity
of factorization.  Thus, at low $q^2$, it is the amount of transverse
polarization that has to be measured with a small relative
uncertainty.  Especially for the semi-leptonic decay such a high
precision measurement of the $q^2$-dependence of the
$D^*$-polarization seems hardly possible at present.  Fortunately,
this seems to be a case where heavy-quark symmetry predictions
receive only minor corrections.  In the infinite quark mass limit one
finds for the ratio of transverse to longitudinal polarization at
some fixed $q^2$
\begin{equation}
\frac{{\rm d}\Gamma_T}{{\rm d}\Gamma_L} =
\frac{4q^2(m_B^2+m_{D^*}^2-q^2)}{(m_B-m_{D^*})^2
\left[(m_B+m_{D^*})^2-q^2\right]}
\label{pol}
\end{equation}
which is subject to QCD- as well as $1/m_Q$-corrections.  The general
structure of the corrections can be found in Ref.~\cite{vcb}.  While
the QCD-corrections can be reliably calculated using perturbation
theory (see e.g. Ref.~\cite{qcd}), the $1/m_Q$-corrections are
model-dependent.  We have calculated the corrections to
Eq.~(\ref{pol}) using the QCD-corrections of Ref.~\cite{qcd} and the
$1/m_Q$-corrections resulting from  a)  an analysis of the wave
function model of Bauer, Stech and Wirbel~\cite{isgur,bsw85} and b) a
sum rule calculation~\cite{sum92}.  Although corrections to
individual form factors in both models are as large as 30\% at
maximum recoil and furthermore vary strongly between both models, the
correction factor to Eq.~(\ref{pol}) in neither one of the two models
deviates from one by more than 5\% (though the deviations in the two
models go in opposite directions).  In Table~\ref{tab:pol}, we
present the prediction for the $D^*$-polarization in semi-leptonic
$B$-decays as a function of $q^2$ obtained from Eq.~(\ref{pol}).  The
quoted errors result from the conservative estimate of a $10\%$
relative uncertainty for $\Gamma_T/\Gamma_L$ at maximum recoil (i.e.\
at $q^2=0$), decreasing linearly to the point of zero recoil, where
the polarization is fixed model-independently.

In nonleptonic $B$-decays, the only polarization measurement
presently available is that of the $D^*$-polarization in the decay
$\bar B^0 \to D^{*+}\, \varrho^-$.
CLEO finds $\Gamma_T/\Gamma_{\rm tot.} = (10\pm 9)\%$~\cite{kh},
which has to be compared with the 12\%
transverse polarization predicted for the semi-leptonic decay at
$q^2=m_{\varrho}^2$ (see Table~\ref{tab:pol}).  In order for this
test to be sensitive to deviations from factorization, the
experimental uncertainty will have to be reduced.

The situation may be more favorable in the decay $\bar B^0 \to
D^{*+}\,D_s^*$ with predicted 48\% of transverse polarization,
hopefully allowing for a measurement with smaller relative
uncertainties.  Also, it will be particularly interesting to see
whether this decay obeys the factorization prediction, as this would
indicate that the factorization assumption may be justified even in
decays with only medium energy release.  However, one should keep in
mind that the QCD coefficient $a_1$ drops out of the ratio
$\Gamma_T/\Gamma_L$, so that from polarization tests alone it will
not be possible to decide whether the short range QCD corrections
represented by the values of $a_1$ and $a_2$ are really independent
of the energy release.  To this end, one would like to test the
validity of Eq.~(\ref{RV}) with $V=D_s^*$ using a value for the decay
constant of the $D_s^*$ as determined independently from a
measurement of the rate for the decay $D_s \to \mu\,\bar\nu$
(employing $f_{D_s}\simeq f_{D_s^*}$, predicted by heavy-quark
symmetry).  In the absence of such a measurement $f_{D_s^*}$ may be
taken from sum rule or lattice calculations.

\acknowledgements
I would like to thank K. Honscheid from the CLEO Collaboration for
drawing my attention to $B$-decays into two vector particles.  Much
of this work is based on a detailed study of exclusive $B$-meson
decays done in collaboration with M.~Neubert, B.~Stech and Q.P.~Xu.

\newpage
\narrowtext

\begin{table}
\caption{Determination of the QCD coefficient $a_1$ from several
nonleptonic $B$ decay modes as a test of the factorization
assumption.  The data are taken from CLEO and ARGUS.  The theoretical
predictions for the branching ratios in the last two rows are those
of Ref.~\protect\cite{buch}}
\begin{tabular}{c|ccc}
Quantity & Experiment & Theory & $a_1$\\
\tableline
$R_{\pi}^*$ [see Eq.~(\ref{RP})] & 1.29$\pm$ 0.22 & $0.97 a_1^2$ &
1.15$\pm $0.10\\
$R_{\varrho}^*$ [see Eq.~(\ref{RV})] & 3.0$\pm$ 0.7 & $2.37 a_1^2$ &
1.13$\pm $0.13\\
$R_{\varrho}^*/R_{\pi}^*$ & $2.5\pm 0.6$ &
$f_{\varrho}^2/f_{\pi}^2=2.4$ & --- \\
BR$(\bar B^0 \to D^+\pi^-)$ & $0.28\pm 0.05$ & $0.214 a_1^2$ &
$1.15\pm 0.10$ \\
BR$(\bar B^0 \to D^+\varrho^-)$ & $0.74\pm 0.22$ & $0.502 a_1^2$ &
$1.21\pm 0.18$ \\
\end{tabular}
\label{tab:ratios}
\end{table}

\begin{table}
\caption{Amount of transverse polarization (in \%) of the $D^*$ in
semi-leptonic $B$-decay.}
\begin{tabular}{c|ccccc}
$q^2$ & 0 & $m_{\varrho}^2$ & $m_{a_1}^2$ & $m_{D_s^*}^2$ & $q_{\rm
max}^2$\\
\tableline
${\rm d}\Gamma_T/{\rm d}\Gamma_{\rm tot.}$ & 0 & 12$\pm$ 1 &
26$\pm$ 2 & 48$\pm$ 1 & $\frac{2}{3}$\\
\end{tabular}
\label{tab:pol}
\end{table}

\end{document}